\documentstyle[10pt,aas2pp4]{article}
\newcommand{\amin}{$^{\prime}$}
\newcommand{\asec}{$^{\prime \prime}$}
\newcommand{\rf}{\reference}
\newcommand{\exvalm}[1]{\mbox{$<$}\mbox{$#1$}\mbox{$>$}}
\newcommand{\um}{~$\mu$m}
\newcommand{\minpoint}{\mbox{$'\mskip-4.7mu.\mskip0.8mu$}}

\newcommand{\gapprox}{\mbox{$\stackrel {>}{_{\sim}}$}}   

\newcommand{\ecs}{\mbox{erg s$^{-1}$ cm$^{-2}$}}
\newcommand{\kmsmpc}{km~s$^{-1}$~Mpc$^{-1}$}

\newcommand{\mc}[2]{\multicolumn{#1}{#2}}
\newcommand{\al}[2]{\mbox{$\alpha_{#1}^{~#2}$}}
\newcommand{\SS}[1]{\S~\ref{#1}}
\lefthead{Rush et al.}
\righthead{Radio Properties of Seyferts}

\begin{document} 



\title{THE RADIO PROPERTIES OF SEYFERT GALAXIES IN THE 12--MICRON AND CFA
SAMPLES}

\author{Brian Rush\altaffilmark{1} and Matthew A. Malkan} \affil{Department of Physics and
Astronomy, University of California, Los Angeles, CA 90095--1562;
rush,malkan@astro.ucla.edu}

\altaffiltext{1}{Current Address: The Carnegie Observatories, 813 Santa Barbara
St., Pasadena, CA 91101--1292}

\and

\author{Richard A. Edelson} \affil{Department of Physics and Astronomy, 203 Van
Allen Hall, University of Iowa, Iowa City, IA 52242;
edelson@spacly.physics.uiowa.edu}

\begin{abstract} 

We report the results of 20, 6, and 2~cm VLA and 1.5~cm OVRO observations of
two similar AGN samples: the optically--selected CfA Seyfert galaxies and the
bolometric--flux--limited 12\um\ active galaxy sample. Every object observed
was detected at 6~cm. Only $\sim6-8\%$ of the 12\um--sample Seyferts (3--4
objects) are radio--loud (and none of the CfA sample), as compared to 15---20\%
for the BQS quasars. These radio-loud objects are compact and have flat
spectra, distinguishing them from the more common radio--quiet objects.

The 6---20~cm slopes of the Seyfert~1s and~2s are similar, with average values
of $\exvalm{\al{6cm}{20}} = -0.66$ and --0.71, respectively. Although several
Seyfert~1s are significantly flatter than this in their 6---20 and/or
1.5---6~cm slopes, there is no systematic trend for either Seyfert type to
display upward or downward spectral curvature.

Excluding the radio--loud quasars, the integrated 6~cm radio luminosity is
linearly proportional to the 60~$\mu{m}$ luminosity over several orders of
magnitude, with on average twice the radio power of normal spirals of the same
far--infrared power. About half of the objects show extended 6~cm emission,
contributing on average 33\% of the total flux. Thus the luminosities of these
extended components alone are comparable to normal spirals of similar infrared
luminosities.

The 12\um\ sample radio luminosity function is slightly higher than that of the
CfA sample. The integrated space density of Seyfert~2s is $\sim2$ times that of
Seyfert~1s over their common range in luminosity. In terms of the standard
unified model, this ratio in space density corresponds to a typical half--angle
of the torus of $\theta\sim48^\circ$.

\end{abstract} 


\section{Introduction} \label{intro} 

The most fundamental division between different types of extragalactic radio
sources is radio--loud vs. radio--quiet, where radio--loud objects are those
with much higher radio luminosities, both in absolute terms
($L_{\mbox{\scriptsize6~cm}}>10^{42}$~erg~s$^{-1}$---Miller, Peacock, \& Mead
1990), and relative to other wavelengths (e.g., a factor of $\sim10^4$ more
luminous at 6~cm for a given [O~III] luminosity---Wilson \& Colbert 1995).
Radio--quiet active galaxies and quasars (generically referred to as AGN
herein) can further be divided into radio-quiet quasars, Seyfert~1s, and
Seyfert~2s (the difference between the first two being, perhaps, just total
absolute luminosity).

Furthermore, quantifying any differences between the average radio properties
of various types (e.g. radio--loud vs. radio--quiet objects or Seyfert~1s vs.
Seyfert~2s) also has direct applications towards unified models which relate
different classes through effects such as relativistic beaming, or
orientation--dependent obscuration. According to the basic unified model for
Seyfert galaxies, the nuclei of Seyfert~2s are intrinsically similar to those
of Seyfert~1s, yet viewed edge--on, so that our view to the inner--most parts
of the nucleus, including the ``broad--line'' region, is obscured by a
molecular torus. More complicated models expand on this picture by including
other parameters which vary from object to object, such as the thickness of the
torus or mass of the central engine. These models can be tested observationally
by the fact that they predict many differences between the multiwavelength
properties of type~1 and type~2 Seyfert galaxies (Antonucci 1993). A general
prediction of this model is that isotropic properties (originating at radii
outside of the torus) will be similar in Seyfert~1s and~2s, but that emission
from the innermost regions will be orientation--dependent and thus will differ
between Seyfert~1s and~2s.

As an example, a potential challenge to the simplest form of these unified
models was found in early studies concluding that type~2 Seyferts have stronger
and larger nuclear radio sources than type~1 Seyferts (de~Bruyn \& Wilson 1978;
Meurs \& Wilson 1984; Ulvestad \& Wilson 1984a,b).  However, these studies were
influenced by selection effects in the Markarian sample, causing the weaker
Seyfert~2 galaxies to be omitted from the samples.  In contrast, samples
selected largely from the CfA redshift survey show no significant difference
between the radio sources in the different Seyfert types (Edelson
1987---hereafter E87; Ulvestad \& Wilson 1989; Giuricin et
al.~1990). Determining which result holds for the true population of Seyferts
in the local universe requires observations of a well--defined sample large
enough for significant statistical analysis.

We have therefore obtained 6~and 20~cm data from the VLA (and some single--dish
1.5~cm fluxes) for two samples of bright, nearby AGN which are mostly
radio--quiet, specifically classified as Seyfert~1s and Seyfert~2s. We use
these data to investigate the characteristics of the observed radio and
multiwavelength properties of Seyferts and to study the differences in these
properties between Seyfert classes.

Since radio--quiet AGN reside in host galaxies that may also contribute
significantly to the overall observed spectral energy distribution, we also
need to determine the relative contribution from the central and extended
components.  With this in mind, we will discuss several of the results
presented herein in terms of a two component model, with the central component
(i.e., the nucleus plus other unresolved flux) accounts for most of the radio
flux, but where an extended component, with a lower radio---infrared flux
ratio, also contributes significantly.

In \SS{obs} (and Appendix~\ref{app:selection}) we discuss the target selection
and observations. The analysis is in \SS{radio-spec}---\ref{rlf}: in
\SS{radio-spec} we discus radio spectral properties; in \SS{compactness}
compactness and extended emission; in \SS{radio-ir} the correlation between
radio and infrared luminosities; in \SS{radio-loud} the frequency of
radio--loud objects in the 12\um\ sample; and in \SS{rlf} radio luminosity
functions. A summary is given in \SS{summary}

\section{Observations and Data Reduction} \label{obs} 


The CfA and 12~$\mu$m Seyfert galaxy samples are believed to be relatively free
of selection effects and systematic biases that have plagued other samples such
as the Markarian Seyferts (see Appendix A; also Huchra \& Burg 1992, RMS93). As
the CfA sample is optically selected, while the 12~$\mu$m sample was selected
in the infrared, they will enable us to compare the radio properties of samples
with very different selection criteria (mid--IR and optical, respectively), as
well as to compare the properties of each sample to those of radio--selected
objects. We therefore obtained VLA 6 and 20~cm data of virtually all objects in
both samples which are observable from the VLA (with the exception of a few
objects which were added to the final definition of these samples after we
began this project).

The new observations presented in this paper were carried out during
1990--1991.  Seven CfA Seyferts with missing VLA data or upper limits in E87
were observed during January~1990. The 12\um--sample Seyferts (that were not
also observed as part of the CfA sample in E87) were observed during two runs
in March~1991.  Single snapshots were taken with integration times of
10~minutes at 6~cm and 2.5~minutes at 20~cm, for 34 objects in each band. Every
object observed was detected above the 3$\sigma$ noise levels of 0.35 at 6~cm,
and most above 1.0~mJy at 20~cm. The observations presented in E87 (both the
VLA and OVRO data), and also discussed here, were taken during July~1983.

To make comparisons at different wavelengths, we combined untapered D--array
measurements at 20~cm, tapered beam data at 6~cm, and single-dish 1.5 cm
observations, to achieve a fairly uniform beamwidth of $\sim$1\minpoint5 FWHM.
In addition, untapered (typical beam FWHM~$\sim15''$) 6~cm measurements are
also used for measuring the flux of the compact region. Since the radio sources
in nearby Seyferts are often partially resolved on longer baselines, this large
beam will yield the most uniform database, without introducing biases by mixing
data from different arrays. The VLA maps were calibrated with the standard AIPS
software, CLEANed with the AIPS task MX, and finally we used the AIPS task
IMFIT to fit to an elliptical Gaussian to each map to measure the central flux
density. The reduction of the OVRO data is described in E87.

Potential variability is not likely to affect our results in any significant
way, since most of our observations are at least quasi--simultaneous. The data
at 6~cm and 20~cm (and, where it exists, at 2~cm) were taken on the same day
for any given object, and the OVRO 1.5~cm observations were made within two
weeks of the VLA observations for the same object (except for NGC~5273, for
which the 20~cm data were taken in 1991 and 1.5~and 6~cm data were taken in
1983).

Table~1 presents the data. Column~1 gives the name of each object, column~2 the
Seyfert type, column~3 the sample (12\um\ and/or CfA), and column~4 the
redshift. The next columns gives the flux densities and uncertainties (in mJy)
at 20~cm, 6~cm (both high (h) and low (l) resolution, for untapered and tapered
beams, respectively), and 1.5~cm. (For the five objects noted in the table
footnote, this last data point is actually a 2~cm observation taken with the
VLA and should be considered only as a lower limit, since the VLA beam size is
much smaller than that at OVRO.) The quoted uncertainty is the quadratic sum of
the statistical errors and an estimated 5\% uncertainty in the
calibration. Upper limits are at 3$\sigma$. The final column indicates where
these data were first reported (R~=~this work; E~=~E87). We list all of our new
observed fluxes and those from E87 in Table~1, so that readers can readily have
available the complete VLA data for both the 12\um\ and CfA samples.

In Table~2 we present derived properties. The first 3 columns are the same as
in Table~1. Columns 4 and 5 give the radio spectral indices, \al{1.5}{6} and
\al{6}{20}, using the tapered 6~cm beam for accurate comparison to the other
wavelengths.  Columns 6 and 7 present the radio--IR spectral indices between
6~cm and both 60\um\ and 12\um. The IRAS data were obtained from Rush, Malkan,
\& Spinoglio (1993---hereafter RMS93) for objects in the 12\um\ sample and from
Edelson, Malkan, \& Rieke (1987) for those CfA--sample Seyferts not in the
12\um\ sample. Column 8 gives the IRAS 25--60\um\ spectral index. Columns 9,
10, and 11 give the 6~cm, 20~cm, and 60\um\ monochromatic
luminosities\footnote{We assume a value of H$_0$~=~75~\kmsmpc throughout this
paper.} ($\nu L_\nu$, in units of $erg$~$s^{-1}$). Column 12 gives the
radio--compactness parameter, defined as $R=S_{6cm,h} / S_{6cm,l}$, following
E87.  We stress that this $R$ parameter is not a ratio of the flux from the
unresolved AGN nucleus to that from the entire galaxy, as others have
used. Rather, it is a ratio of the ``central'' flux to that from the entire
galaxy, where by ``central'' we mean the less--extended flux in a general
sense. This central flux is not to be confused with the unresolved ``nuclear''
flux, since it would include this flux plus any double, triple, and/or
jet--like sources associated with the nucleus, as well as circumnuclear or
inner disk emission related to a starburst. With this in mind, we will refer to
the numerator and denominator of this $R$ parameter (i.e., the $S_{6cm,h}$ and
$S_{6cm,l}$ values) as the ``central'' and ``total'' 6~cm flux (where total =
central plus ``extended'') throughout this paper. The usefulness of this $R$
parameter as we have defined it is very similar to, but less powerful than a
ratio of nuclear--to--extended flux. A plot of $R$ versus nuclear--to--extended
flux for a sample of Seyferts flux should be monitonically increasing in
general, with both ratios being larger in more nuclear--dominated
sources. Thus, $R$ gives some indication of the extent to which an object is
nuclear--dominated, i.e. how ``compact'' it is, hence the name ``compactness
parameter.''

\section{Radio Spectral Properties} \label{radio-spec} 

We have measured the 6---20~cm spectral slope and, when possible, the
1.5---6~cm slope. The average value of \al{6}{20} is $-$0.66$\pm$0.04 for
Seyfert~1s and $-$0.71$\pm$0.04 for Seyfert~2s (all uncertainties quoted herein
represent one standard deviation of the mean unless otherwise noted). For those
objects with 1.5 or 2~cm observations, we have plotted \al{6}{20} versus
\al{1.5}{6} in Figure~1. (All spectral slopes referred to here are such that
$S_{\nu} \propto \nu^{\alpha}$.) Symbols are explained in the caption. For the
few objects for which we have 2~cm D--array data instead of 1.5~cm OVRO
single--dish data, we have assumed \al{1.5}{6}~=~\al{2}{6}\footnote{These
objects usually have very low fluxes at 2~cm, indicating less high--frequency
flux than the other galaxies in this plot. This is most likely due to the fact
that the VLA in D--array has the greatest resolution at the shorter wavelengths
and has resolved these objects, whereas the OVRO single--dish flux is from the
entire galaxy.  Thus, the 2~cm point is considered a lower limit, and thus the
points in the plot are marked as lower limits to \al{1.5}{6} (right--pointing
arrows)}. This figure is the same as Figure~3 in E87, with the addition of the
points at 2~cm, with the upper/lower limits displayed, and with changes in some
object types due to better, more recent optical spectra.

We note that one tentative result suggested in E87 cannot be confirmed with
this data, namely that Seyfert~1s are more likely than Seyfert~2s to have a
high--frequency excess, i.e. spectral curvature. This can be seen by noting the
solid line, which represents \al{1.5}{6}~=~\al{6}{20}. Most objects (for which
both slopes are detections) are within 0.1--0.2 of this line. Although this
distance is larger than the typical intrinsic uncertainty in our measured
slopes, it is on the order of, or smaller than, the typical uncertainties of
the slope measurements which result from measuring the flux from less than the
entire galaxy (and even from different regions at each wavelength). This point
is illustrated in Figure~2, where we the compare our 6---20~cm (VLA) slopes
with those measured for the whole galaxy from the single--dish fluxes in the
northern sky survey (Becker, White, \& Edwards 1991; White \& Becker 1992;
shown for all objects in both samples). As can be seen, the differences in
slope are typically 0.1--0.3. Such differences would probably not cause any
{\it systematic\/} change in Figure~1 if whole--galaxy fluxes were used, as is
implied by the absence of any correlation in this plot between the slope
difference and our slopes. Neither is there any trend with redshift, as one
might expect if there is a strong radio color gradient in these galaxies, which
could cause the slope difference to be stronger in nearby sources. (A similar
result, of the same or smaller magnitude, is found when we compare our
1.5--6~cm slopes with those formed by combining our 1.5~cm single--dish fluxes
with the Becker et al. 6~cm fluxes.)

Finally, we point out that there is one systematic effect apparent in Figure~1,
namely that those few objects with at least one or both slopes being very flat
are all Seyfert~1s. However, this does not comment on spectral curvature,
because either slope can be the flat one.  The dotted lines in Figure~1 enclose
those objects in the lower left with both slopes steeper than --0.6. All 8
objects outside of this box with detections in both axes are Seyfert~1s.  The
three objects with one or the other slope being {\it very\/} flat (i.e.,
outside of the box by twice the typical uncertainty discussed above), which
would not simply result from measurement differences, are individually
labeled. That the other 5 flat objects are all Seyfert~1s is probably also
physically meaningful, since it is not likely that this would happen by chance,
and since there is no general tendency (as would be seen in Figure~2) for
Seyfert~1s to be flatter in our measurements than in the single--dish
measurements.

\section{Extended Emission} \label{compactness} 

To investigate the relation between the radio--compactness parameter, $R$, and
other radio properties of Seyferts, we have plotted L$_{6cm}$ versus $R$ in
Figure~3. Here we see that the few radio--loudest objects are very compact, all
having $R\sim1$ ($R$ is typically accurate to $\pm10\%$---less for the fainter
objects---hence the few values of $R$ greater than~1).  We note that the more
luminous and compact objects in our sample are also among those with the
flattest spectral slopes.  These facts are consistent with models in which the
radio--loud objects are compact, flat spectrum radio sources and that they have
type--1 Seyfert nuclei with the radio emission directed towards our line of
sight. The most luminous Seyfert~2s, on the other hand, have steeper spectra,
but are still compact at our D--array resolution (they may, however, be shown
to have less compact cores if observed with higher resolution). Three
Seyfert~1s and no Seyfert~2s have $L_{6cm}>10^{40}$, have the flattest spectra
($\al{6}{20}>-0.2$) and are very compact ($R=1$).

Thus, we see a clear distinction in properties ($L_{6cm}$, \al{6}{20}, and $R$)
between the very few radio--loud objects and the more numerous, relatively
radio--weak objects in our sample (see \SS{radio-ir} for further discussion of
radio--loud versus radio--quiet objects in our sample). For the majority of the
(radio--weak) galaxies, a slight trend in the same sense is found, with much
scatter, indicating that some of these objects may be harboring very weak
compact cores. This is shown by the straight line in Figure~3, which is the
best--fit line to all objects in the plot (with slope~=~2.41 and r~=~0.23), and
the dashed line which excludes the 7 most luminous (labeled) galaxies (with
slope~=~1.56 and r~=~0.17; excluding only the two strictly radio--loud objects
would make the slope even steeper). (See below for explanation of the curved
lines in this figure.)

We find that, although L$_{6cm}$ is correlated with distance in our sample (as
expected; r~=~0.68), $R$ is not (r~=~0.06), implying that the correlation
between L$_{6cm}$ and $R$ is not simply an artifact of redshift. This also
implies that the variation in $R$ in our sample represents the {\it
intrinsic\/} range of extended 6~cm emission among Seyfert galaxies. Thus, we
find in both samples that both Seyfert~1s and~2s have steep radio spectra with
resolved structure on the \gapprox1\amin scale. This suggests that the low
frequency and low resolution emission may be dominated by optically thin
synchrotron emission from an optically thin source, such as the galactic disk.

The average value of $R$ for all galaxies is $\sim$0.83, with no difference
between Seyfert type or between the two samples. Roughly half of the objects
have R$<$0.9, i.e. extended 6~cm emission is found in about half of the
observed galaxies, with an average value for $R$ among those objects of
$\sim0.67$ (i.e., an extended component contributes about $0.33\%\pm0.17\%$ to
the total flux).  This represents a significant contribution from the
underlying galaxy, which must be taken into account when considering
measurements such as total luminosities where both the central and extended
components are significant, as well as spectral slopes and flux ratios where
the two combined components may have different values.

Accordingly, we have also plotted in Figure~3 several curved lines representing
a simple physical model. In this model, the total 6~cm luminosity is the sum of
the central and extended components (L$_{cen}$ and L$_{ext}$, respectively),
and $R$ is the ratio of central to total (i.e., central to
central--plus--extended) luminosity. (Note that the extended component is a
lower limit to the luminosity of the disk of the galaxy, as the latter will
also emit at least some flux at radii within the ``central'' component.)  Each
curve starts at a given value of $\log{L_{ext}}$ (36.5, 37.5, and 38.5 for the
lower, middle, and upper curves, respectively) at $R=0$, and increases with $R$
(i.e., as the fraction of central luminosity increases). Although the scatter
of the data is quite large, the general shape of these curves matches the data:
luminosity is slightly correlated with compactness for small values of $R$, but
increases sharply at the highest values of $R$. The curve representing an
extended--component luminosity of $\log{L_{ext}}=37.5$ goes right through the
center of the data (and roughly also the best--fit lines), but values an order
of magnitude higher or lower than this are required to reproduce all the
points.

\section{The Radio--Infrared Correlation} \label{radio-ir} 


Figure~4 shows a plot of the 6~cm versus 60\um\ monochromatic luminosities. The
results of a bivariate regression to all the data (excluding the superluminal
quasar 3C~120 and the BL~Lac object OJ~287---discussed further in
\SS{radio-loud}) is $L_{6cm}\propto L_{60\mu m}^{1.05}$, indistinguishable from
a linear correlation. We therefore show, with dotted line \#1 the best fit
obtained when the slope is constrained to~1, yielding the relation
$L_{6cm}=10^{-b}\times L_{60\mu m}^1$, with $b=5.31$. The linear
proportionality for normal galaxies (determined by Bicay \& Helou 1990, using
their 20~cm---60\um\ relation and assuming a value of \al{6}{20}=$-$0.7),
which has the value $b=5.64$, is shown with line \#2. For comparison, we also
show the lines from RMS93, representing their fit to Seyferts (\#3; $b=5.0$)
and non--Seyferts (\#4; $b=5.61$). Those fits were done using ``survival
analysis" procedures with the ASURV software package (La~Valley, Isobe, \&
Feigelson 1992) to account for many upper limits in the radio fluxes,
particularly of Seyferts.  (Such procedures assume that both the detected and
undetected objects were drawn from the same homogeneous sample, which can
explain why the fit to Seyferts from RMS93 is higher than that of this work.)

Comparing these lines shows Seyfert galaxies to have excess 6~cm emission
relative to that at 60\um, as compared to non--Seyferts, by about a factor of
two. There is no significant difference in this relation between Seyfert~1s
and~2s. This can be explained if Seyferts are {\it not\/} like normal spirals,
but include a mix of ``central'' radio--plus--IR light with galaxy
radio--plus--IR light. In this scenario, the central component, more dominated
by the Seyfert nucleus is the component with the higher radio---IR flux
ratio. To explain this, we have calculated curves similar to those in
Figure~3. As in that case, we model the luminosity at 6~cm as the sum of
central and extended components, and now we do the same for the 60\um\
luminosity.\footnote{We stress that this is a simplified model and that the
actual case is probably more complicated (for example, the radio---infrared
ratio of the central component may increase with luminosity). However, such
modeling will help to illustrate the extent to which the emission from Seyfert
galaxies can be explained as resulting from the sum of a central plus an
extended component, each with different values of the ratio of radio to
infrared luminosity.} For the 6~cm---60\um\ slope for the extended component,
we use the value derived for normal galaxies by Bicay \& Helou (1990; $b=5.64$,
as mentioned above). The different model curves represent the following
variations of parameters: each of two {\it sets\/} of model curves starts at a
locus (denoted by an open star), which corresponds to a given value of the
extended--component luminosity at 6~cm ($L_{ext}=10^{36.5}$, and $10^{38.0}$,
for the lower left and upper right sets, respectively). Four curves then span
out from each star, corresponding to different color {\it central\/} components
($b=5.64$, 5.31, 5.0, and 4.5, representing normal galaxy--like, Seyfert--like,
radio--strong, and very radio--strong colors; as labeled on the lower set of
curves). Finally, along each of these curves, $R$ varies from 0.01 at the star
up to 0.999 at the end of the curve (with transverse lines drawn at $R=0.50$
and $R=0.90$).

As defined, when $b_{cen}=b_{ext}=5.64$, the curves follow the line from Bicay
\& Helou (1990) for normal galaxies, while when $b_{cen}=5.31$, 5.0, or 4.5,
the curves start at the normal galaxy line (at $R=0$) and asymptotically
approach the Seyfert lines (as $R\rightarrow1$). We see that the different sets
of curves are highly degenerate. For example, a very compact object with
$L_{ext}\sim10^{36.5}$ may have a similar total luminosity to a less compact
object with $L_{ext}\sim10^{38}$. Furthermore, the fact that the scatter is
much larger than the observational uncertainties in the data indicates that,
some objects would have to have much higher or lower 6~cm---60\um\ flux
ratios than the values used in this simplified model.

We apply this model further in Figure~5, where we plot the 25---60\um\ infrared
slope versus the radio compactness parameter, $R$. We chose this infrared slope
because it is often used to select for ``warm" IRAS galaxies which are often
Seyferts (e.g., Low et al. 1988). Here we also assume the constant
6~cm---60\um\ ratio from Bicay \& Helou for the extended component. The
different curves represent various values of this ratio for the central
component. The highest (solid) curve assumes $b_{cen}=b_{ext}=5.64$, the next
highest one $b_{cen}=5.31$, and the two dotted ones $b_{cen}=5.0$ and 4.5,
respectively. The two horizontal lines represent the values of the 25---60\um\
ratio assumed for the extended and central commponents (typical values for
normal galaxies and for quasars in our 12\um\ sample, respectively). The curves
all connect the lower horizontal line at $R=0$ to the upper horizontal line at
$R=1$ (i.e., the values chosen for \al{25}{60} simply determine the start and
end points of each curve, without affecting their shape). These curves show us
that a very radio--strong central component is necessary to match some of the
data points given this model, while a range in the radio---infrared ratio of
the central component is still necessary to match all the data. Alternatively,
a cooler far--IR slope of the central component (i.e., still significantly
hotter than the extended component, but not by as much as assumed in the model
plotted), again combined with a wide range in radio strength, could also
produce a family of curves which span most of the data points without requiring
such extreme high values of radio strength (i.e., $b\sim5$ instead of
$b\sim4.5$ would be sufficient).


\section{Radio--Loud Objects in the 12--Micron Sample} \label{radio-loud} 

From these figures we see that a few objects in the 12\um\ sample can clearly
be distinguished from the rest as being radio--loud. For comparison, Kellerman
et al. (1989) observed the BQS quasars at 6~cm, finding about 15---20\% of the
that sample to be radio--loud, having $\log F_{\nu,rad}/F_{\nu,opt}$ from 1.5
to 3, while the rest have values around --1 to 1.5. These correspond to values
for $\log(L_{6cm}/ L_{60\mu{m}})$ of --4.3 to --1.8 for radio--loud objects and
--5.8 to --4.3 for radio--quiet objects (assuming a typical optical---60\um\
conversion for the BQS quasars---Spinoglio et al 1995). From Figure~4 we see
that $\log(L_{6cm}/ L_{60\mu{m}})$ is --2.6 for both 3C~120 and OJ~287. We did
not observe 3C~273 at 6~cm due to scheduling difficulties, but using the 6~cm
flux of 34.9~Jy from Kuehr et al. 1981 yields $\log(L_{6cm}/
L_{60\mu{m}})=-2.3$, making this the radio--loudest object in the 12\um\
sample.  Thus, we see that only $\sim6-8\%$ of the 12\um\ sample is radio--loud
(3 of $\sim$50 observed objects; 4 including Mkn~463 which is borderline,
having $\log(L_{6cm}/ L_{60\mu{m}})=-4.3$), which is significantly less than
the 15---20\% found for the optically--selected BQS quasars, the difference
probably being a function of redshift (and the CfA sample includes {\it no\/}
radio--loud objects, with the radio--loudest objects of that sample being
Mkn~231 with $\log(L_{6cm}/ L_{60\mu{m}})=-4.5$). Furthermore, there is a
bimodal distribution of radio--loudness, in that these three objects exceed all
the others by a factor of \gapprox~100 in radio--loudness. Although these
objects also have flat slopes and compact radio emission, such properties are
also observed in a few other objects e.g., Mkn~231). Therefore, the most
clearly distinguishing trait of these objects is their high radio luminosity as
compared to that at other wavelengths. In each of the three cases where we see
this in our sample, the radio emission is thought to be anisotropic, and beamed
preferentially (but not necessarily directly) towards us.

\section{Radio Luminosity Functions} \label{rlf} 

We have constructed radio luminosity functions (RLFs) at 6~cm for both the
12\um\ and CfA samples, for individual Seyfert types and for all Seyferts, in
order to determine the true RLF for Seyfert galaxies in the local universe.
These RLFs have been derived using the $V/V_{max}$ method (Schmidt 1968;
Schmidt \& Green 1983), $$\Phi={4\pi\over \Omega{\it f\/}\Delta{L}}\sum{1\over
V_{max}},$$ where $V_{max}$ was individually computed for each galaxy in the
sample. We followed the method of E87 for calculating the luminosity function
of a sample at a wavelength other than the wavelength at which the sample was
defined. We thus use $$V_{max} = \mbox{min} (V_{max,survey} V_{max,radio}),$$
which represents the maximum volume of space accessible by an object detected
at the survey wavelength (mid--IR and optical for the 12\um\ and CfA sample,
respectively) and at radio wavelengths. This is equivalent to deriving the RLF
from the IR (or optical) luminosity function and from the bivariate radio--IR
(or radio--optical) luminosity distribution function (Elvis et al. 1978; Meurs
\& Wilson 1984). We stress, however, that these (and all other) bivariate
luminosity functions can only be considered as lower limits to the true space
density of Seyferts. This is because the most extreme objects (i.e., those with
optical/IR fluxes below the optical/IR survey limit, yet radio fluxes above the
radio detection limit) will be excluded, having not been included in the sample
in the first place, even though they would have been detected in the radio. We
used a 6~cm flux limit of 0.35~mJy, (representing a typical 3$\sigma$ noise
level of the 20~cm maps), a 12\um\ flux limit of 0.30~Jy (corresponding to the
survey limit of the 12\um\ sample), and an optical flux limit of 6.25~mJy at
4500A (corresponding to the magnitude limit of $m_{pg}=14.5$ of the CfA
sample---Huchra et al. 1992). The fraction $f$ represents that fraction of the
objects in the sample which were observed.\footnote{For the 12\um\ sample, 7
objects were not observed because they were too far south to be reached from
the VLA, 3 more were not observed because of scheduling limitations, and 13
because they were not known to be Seyferts at the time of SM89 and were not
included in our sample. Thus, $f=(71-23)/71=0.68$. Similarly, for the CfA
sample, only one object was not observed due to scheduling constraints and one
was not originally known to be a Seyfert, thus $f=(49-2)/49=0.96$.} For bins of
width 0.4 in $\log L$ and a RLF proportional to Mag$^{-1}$, $\Delta{L}=1$.  The
errorbars represent the 90\% confidence interval, based on Poisson statistics,
calculated using the equations from Gehrels (1986) which are accurate for even
very small numbers of data points.

Figure~6 shows the radio luminosity function for both the 12\um\ and CfA
samples of Seyferts (all types combined). Tables~3 and~4 tabulate the values of
the RLF for all Seyferts (and as separated into~1s and~2s) in the 12\um\ sample
and the CfA sample, respectively. We fit each RLF to a single power--law
(straight lines), with the results plotted on the graph (in each case the
points are weighted by the number of objects they represent, hence the lines
look higher than would ones which weigh each point evenly). Both RLFs are
fitted well by a power--law (solid line with $r=-0.95$ for the 12\um\ RLF and
the dotted line with $r=-0.91$ for the CfA RLF). The 12\um\ RLF has a steeper
slope ($-$1.01, vs. $-$0.72 for the CfA sample). The integrated RLF for the
entire 12\um\ sample is higher than that for the CfA sample, primarily because
it is higher at low luminosities (similar results are obtained for 20~cm
RLFs). This, as well as the flatter slope of the CfA RLF could result from
low--luminosity Seyfert~1s being underrepresented in the CfA sample when the
weak, broad components of their emission lines are diluted beyond recognition
in the more distant objects (Persic et al. 1989; Huchra \& Burg 1992).  We have
also shown for comparison the RLF from Ulvestad \& Wilson (1989), denoted by
x's and a dotted line. The RLF of this distance--limited sample agrees with the
others above $\log L=38$, but is slightly lower below this level.  It appears
much lower at the very lowest luminosities, but these points are less
meaningful, since they only represent 1--2 objects per bin. Furthermore, part
of the difference is caused by the fact that the Ulvestad \& Wilson RLF is
based on fluxes measured in the A and A/B arrays, which represent a smaller
area of each galaxy, shifting their RLF to the left as compared to our D--array
RLF.

Figure~7 shows the RLFs for individual Seyfert types (1s and~2s) in the 12\um\
sample. The RLF of Seyfert~1s extends with a similar power--law slope to very
high luminosities, while the Seyfert~2s RLF has a sharp high--luminosity
cutoff, reminiscent of the cutoff above the $L_{\star}$ knee in the optical
luminosity function of normal galaxies. Over most luminosities ($\log L>37.4$),
where we can accurately measure the RLFs of both Seyfert types, we find the
space density of Seyfert~2s to be $\sim$2 times that of Seyfert~1s (i.e., about
1/3 are Seyfert~1s), although Seyfert~1s extend to higher luminosities (similar
to the far--infrared luminosity functions calculated in RMS). This has
implications for the unified model in that, if the 20~cm emission is isotropic,
then there are $\sim$2 objects observed to be Seyfert~2s for each intrinsically
similar Seyfert~1. In the context of this (very simplified) model, with
orientation to our line--of--sight being the primary factor distinguishing
Seyfert~1s from~2s, this ratio in space density corresponds to a typical half
opening angle of the torus (within which an object would be observed as a
Seyfert~1) of $\theta \sim\cos^{-1}(1-0.33) \sim48^\circ$. We also note that
the radio--loud objects in the 12\um\ sample account for only 0.04\% of the
integrated luminosity function at 6~cm, however this is only a lower limit as
such objects are the most likely ones to be missed when calculating a bivariate
luminosity function.

A similar plot is shown in Figure~8 for individual Seyfert types for the CfA
sample. We see here and in Table~4 that the space density of Seyfert~2s in the
CfA sample is even less than the Seyfert~1s ($\sim$0.8 times as many
Seyfert~2s), probably due to the fact that the CfA sample, being selected at
optical wavelengths, is biased against heavily reddened Seyfert~2s which have
had much of their optical flux reprocessed into the far--infrared. This may
also explain why the 60\um\ luminosity function of the 12\um\ sample was found
to be higher than that of the CfA sample for both Seyfert~1s and Seyfert~2s
(RMS93).

\section{Summary and Conclusions} \label{summary} 

We have used the VLA in the compact D--array to obtain nearly complete 6~and
20~cm observations for the mid--IR selected 12\um\ Seyfert Galaxy sample and
the optically--selected CfA Seyfert Galaxy sample. We also have analyzed (from
E87) 1.5~cm OVRO data for the CfA sample. The main results are as follows:

There is no significant difference in the {\it average\/} 6---20~cm slopes
between Seyfert~1s and~2s (\al{6cm}{20}$\sim$0.7), consistent with the standard
unified model. There is no systematic trend for either Seyfert type to display
upward or downward curvature, but a few Seyfert~1s have particularly flat
6---20 or 1.5---6~cm slopes.

We have calculated a simple model in which the spatial distribution of the
radio and infrared emission from Seyferts comes from two components: (1) an
extended/disk component which has the same ratio of radio---infrared flux and a
similar luminosity as normal spirals; and (2) a central component with emits
relatively more radio luminosity for a given infrared luminosity. The central
component contributes significantly to the radio--IR emission from Seyferts,
but is much less dominant in normal spirals.

Calculations based on this model describe the following properties of our data:
(1) about half of the galaxies have extended emission at 6~cm, which
contributes an average of $\sim$33\% to their total flux; (2) Seyferts are
shown to have excess 6~cm emission relative to non--Seyferts of similar far--IR
luminosity, by about a factor of two; and (3) among Seyferts, the 6~cm and
60\um\ luminosities are linearly proportional over more than the 3 orders of
magnitude spanned by our data.

Three objects in the 12\um\ sample (and none in the CfA sample) are clearly
radio--loud, and have extreme properties as compared to the rest of the sample.
These objects are the most luminous, have the strongest radio---IR flux ratios,
are compact ($R=1$), and have the flattest spectra (\al{6cm}{20cm}~$\sim$~0).
Thus, there is a clear distinction between these few radio--loud objects and
the radio--quiet objects which dominate these samples.  The fraction of
radio--loud objects is significantly less ($\sim6\%$) than in other,
higher--redshift samples, such as the BQS quasars.

Radio luminosity functions were derived for both the 12\um\ and CfA samples.
Both samples' RLFs are fitted well by a single power-law. The 12\um\ RLF is
slightly higher, especially at low luminosities. The space density of
Seyfert~2s in the 12\um\ sample is about 2 times that of Seyfert~1s over their
common range in luminosity, but the RLF of Seyfert~1s extends to higher
luminosities. In terms of the standard unified model, this ratio in space
density corresponds to a typical (half) opening angle of the torus (within
which an object would be observed as a Seyfert~1) of $\theta\sim48^\circ$.

\acknowledgements

We thank the VLA TAC for providing us with the telescope time during programs
AE63 and AE76, and the VLA AOC and OVRO personnel who helped us with the data
reduction. This work was supported in part by NASA grant NAG~5--1358.  This
research has made use of data obtained through the High Energy Astrophysics
Science Archive Research Center Online Service, provided by the NASA-Goddard
Space Flight Center.

\appendix

\section{Target Selection and Classification} \label{app:selection} 

\subsection{The 12 Micron and CfA Samples} \label{app:selection_samples} 

We chose to define our original sample of galaxies from the IRAS {\it Point
Source Catalog, Ver.~2\/}, flux limited at 12\um\ (Spinoglio \& Malkan 1989;
hereafter SM89), since that is the IRAS wavelength which most strongly selects
for the hot continua universally produced by active nuclei (whether they are
thermal or nonthermal) and is long enough to reject nearly all the flux
produced by stars in the host galaxy. This original 12\um\ sample contains the
390 galaxies above a flux limit of 0.30~Jy, with $|b|\geq25^\circ$ (to avoid
galactic contamination), as well as $F_{60\mu m} \geq F_{12\mu m}$ and/or
$F_{100\mu m} \geq F_{12\mu m}$ (to select galaxies instead of galactic
objects).  This sample is not only complete down to a 12\um\ flux limit, but
also with respect to {\it bolometric\/} flux of $2\times10^{-10}$\ecs\ (RMS).
This sample, as reported in SM89, contained 59 galaxies {\it known\/} to harbor
Seyfert nuclei. Forty--two of these Seyferts are observable from the VLA and it
is for these objects that we have obtained 6~and 20~cm D--array observations.
(It is now known that several other objects in the original 12\um\ sample are
Seyferts, but were not identified as such at the time of SM89. These objects
are properly identified in the Extended 12\um\ Sample---RMS93; discussed in
Appendix~\ref{app:selection_extended}).

We have also observed the Seyfert galaxies in CfA Galaxy sample which is
complete down to an optical flux limit of $m_{Zw}=14.5$ (Huchra \& Burg 1992).
VLA data were presented in Edelson 1987 for 42 of the 50 CfA--sample Seyferts.
We have obtained 6~and 20~cm observations for 6 CfA Seyferts which were added
to the sample after that time, as well as several 2~cm fluxes to compare to the
single--dish 1.5~cm observations from OVRO in E87 (the 2~cm fluxes being made
at much higher resolution are used only to estimate lower limits to the 1.5~cm
fluxes). Four galaxies were added to the final definition of the CfA sample
(Huchra \& Burg 1992; Osterbrock \& Martel 1993) after we began this work and
thus we don't have VLA observations for these objects.

\subsection{Object Classification} \label{app:selection_class} 

In both Tables~1 and~2, classification into type~1.0, 1.5, 1.8, 1.9, and~2 for
all galaxies in the CfA sample (including those objects which overlap with the
12\um\ sample) is from Osterbrock \& Martel (1993), who compiled a consensus
from their own observations and several other works (e.g., Huchra \& Burg 1992
and Dahari \& De~Robertis 1988) based on optical spectrophotometry. For most
galaxies in the 12\um\ sample only, detailed classification into Seyfert
sub--types is not yet available, and thus we have noted the classification
simply as type~1 or~2, based on references in the literature and popular
catalogs (e.g., Hewitt \& Burbidge 1989, 1991; Ver\'on--Cetty \& Ver\'on 1991),
as well as on some of our own spectra. (However, a work is in progress---Rush,
Malkan, \& Spinoglio 1996---in which we will examine high signal--to--noise
spectrophotometry for all Seyferts in the Extended 12\um\ sample to determine
precisely their Seyfert sub--class.) We note that this may slightly skew those
results which focus on differences between Seyfert~1s and~2s (e.g., the
luminosity functions), as it is likely that a handful of 12\um--sample objects
which we now consider to be Seyfert~2s are actually Seyfert~1.8--1.9s, and thus
should be considered Seyfert~1s when dividing the objects into only two
classes.  (Even though the spectra of a Seyfert~1.8---1.9 looks more like that
of a Seyfert~2 than a Seyfert~1, we consider them to be Seyfert~1s when using
only two classes. This is because the detection of slight broad wings to the
optical emission lines indicates the presence of a directly observable
broad--line region, which {\it physically\/} defines a Seyfert~1.8---1.9s as
being Seyfert~1s---see, e.g., Goodrich 1989, 1990.)

\subsection{The Complete List of 12--Micron--Sample Seyferts}
\label{app:selection_extended} 

This paper has studied the radio properties of the CfA Seyfert Galaxy sample
and of the {\it original\/} 12\um\ Seyfert Galaxy sample. For completeness, we
here mention the {\it Extended\/} 12\um\ sample (RMS93) and compare its
contents to those of the original 12\um\ sample.

As mentioned in \SS{app:selection_samples}, the original 12\um\ sample was
selected from the IRAS {\it Point Source Catalog, Ver.~2\/} with a 12\um\ flux
limit of 0.30~Jy (SM89). To probe lower fluxes, we selected candidates for the
Extended 12\um\ sample from the IRAS {\it Faint Source Catalog, Ver.~2\/} and
then defined the sample as those galaxies having SCANPI/ADDSCAN whole--galaxy
12\um\ fluxes above 0.22~Jy. By using the FSC--2, which is complete to a lower
flux limit than the PSC--2, the extended sample contains over twice as many
(893) galaxies.

The original 12\um\ sample contained 58 galaxies {\it known\/} to harbor
Seyfert nuclei at the time of SM89, and is now known to contain at least 71
Seyferts.  Similarly, the Extended 12\um\ sample includes 122 known Seyferts,
and it is likely that several galaxies in this sample have yet to be identified
as Seyferts. Such objects are more likely to be Seyfert~1.8s, 1.9s and~2s which
are often harder to identify. They are also likely to be found in those
positions in multiwavelength parameter space which are usually occupied by
Seyferts.

\section{Comparison of Compactness Parameter With Other Samples}
\label{app:comparison} 

We have checked the accuracy of our measured fluxes by comparing our results to
those from the 4.85~GHz northern sky survey of Becker et al.  (1991), which has
40\asec pixel size and 3\minpoint5 angular resolution. Figure~9 shows a graph
of the ratio of our tapered 6~cm flux to the flux from that catalog versus our
6~cm compactness parameter, $R$. We find, in general, that both values are near
one for the majority of objects (meaning that all three fluxes---S$_{6cm,l}$,
S$_{6cm,h}$, and S$_{4.85GHz}$---are roughly equal), while those galaxies with
values of $R$ much less than one also have low values of the other ratio,
implying that they are simply the most extended. That most values are near 1
further implies that there is little evidence for variability between these
observations.

We have also compared our 6~cm compactness parameter to the 2295~MHz flux
density from Roy et al. (1994) in Figure~10 (upper limits to the 2295~MHz flux
represent 5 times the rms noise in the fringe-frequency spectrum). We see that
the detections are mostly of our most compact objects and the non--detections
are spread over all values of R. This is as one would expect if the extended
sources are resolved by the 275~km interferometer and thus are less likely to
be detected. In fact, only six objects with R~$<$~0.85 were detected at 10~cm
and are individually labeled in Figure~10. Three of these objects (NGC~1068,
MKN~841, and TOL~1238-364) are among the 10~cm--brightest objects in this plot
and thus one would expect them to be easily detectable. Our data also shows
NGC~1365 to be one of the brighter radio sources.

However, we do not see the trend claimed in Roy et al. (1994), that compact
radio structures are much more common in Seyfert~2s than in Seyfert~1s. This
can be explained by the fact that they note only the {\it combined\/}
statistics of two optically selected samples (the CfA sample from E87 and the
sample from Norris et al. 1990) and the the 12\um\ sample. When examined
individually, we find that there is {\it no\/} significant difference between
the detection rates of Seyfert~2s and Seyfert~1s in {\it any one\/} of these
samples alone. However, the overall detection rate is different in each of
these three samples (about 65\%, 50\%, and 20\% in the 12\um, CfA, and Norris
et al. samples, respectively). The Norris et al. sample has the lowest
detection rate of {\it both\/} Seyfert~1s and Seyfert~2s (5~of~28 and 2~of~6,
respectively), which is likely due to the fact that it is a higher--redshift,
fainter sample. This sample also has the most Seyfert~1s and fewest Seyfert~2s
observed of the three samples. Thus, when these different samples are averaged
together, the low detection rate of the Norris et al. sample artificially drags
down the combined 3--sample detection rate of Seyfert~1s far more than the rate
for Seyfert~2s.

\clearpage \onecolumn
 
\begin{table} 
\begin{center} 
TABLE 3\\
\vspace{0.4cm}
{\sc Radio Luminosity Function for the 12--Micron sample}\\
\vspace{0.3cm}
\footnotesize
\begin{tabular}{ccccccc}\hline\hline
\mc{1}{c}{}&
\mc{2}{c}{Seyfert 1s}&
\mc{2}{c}{Seyfert 2s}&
\mc{2}{c}{Both Types}\\ \cline{2-3}\cline{4-5}\cline{6-7}
\mc{1}{c}{$\log L_{rad}^a$}&
\mc{1}{c}{$\log\Phi$}&
\mc{1}{c}{}&
\mc{1}{c}{$\log\Phi$}&
\mc{1}{c}{}&
\mc{1}{c}{$\log\Phi$}&
\mc{1}{c}{}\\
\mc{1}{c}{(ergs~s$^{-1}$)}&
\mc{1}{c}{(Mpc$^{-3}$ M$^{-1}$)}&
\mc{1}{c}{$N$}&
\mc{1}{c}{(Mpc$^{-3}$ M$^{-1}$)}&
\mc{1}{c}{$N$}&
\mc{1}{c}{(Mpc$^{-3}$ M$^{-1}$)}&
\mc{1}{c}{$N$}\\
\hline
 36.40 & $-$3.70 &   1 &  ...    & ... & $-$3.70 &   1\\ 
 36.80 & $-$4.08 &   1 & $-$5.44 &   1 & $-$4.06 &   2\\ 
 37.20 & $-$4.00 &   2 & $-$3.82 &   3 & $-$3.60 &   5\\ 
 37.60 & $-$5.19 &   1 & $-$4.21 &   2 & $-$4.17 &   3\\ 
 38.00 & $-$6.22 &   1 & $-$4.62 &   4 & $-$4.61 &   5\\ 
 38.40 & $-$4.61 &   6 & $-$4.60 &   6 & $-$4.30 &  12\\ 
 38.80 & $-$5.93 &   5 & $-$5.56 &   2 & $-$5.40 &   7\\ 
 39.20 & $-$6.85 &   1 & $-$5.78 &   6 & $-$5.74 &   7\\ 
 39.60 &  ...    & ... & $-$6.97 &   1 & $-$6.97 &   1\\ 
 40.00 &  ...    & ... & $-$6.26 &   1 & $-$6.26 &   1\\ 
 40.40 &  ...    & ... & $-$7.40 &   1 & $-$7.40 &   1\\ 
 40.80 & $-$7.15 &   1 &  ...    & ... & $-$7.15 &   1\\ 
 41.60 & $-$6.82 &   1 &  ...    & ... & $-$6.82 &   1\\ 
 43.20 & $-$9.47 &   1 &  ...    & ... & $-$9.47 &   1\\ 
\hline
\end{tabular}
\renewcommand{\baselinestretch}{1}
\parbox[c]{12cm}{\noindent\footnotesize
\hspace*{.3cm}$^a$~Central luminosity of a bin 0.4 units wide in $\log L$, which
is equivalent to a width of 1 magnitude.}
\end{center}
\end{table}

\begin{table}
\begin{center}
TABLE 4\\
\vspace{0.4cm}
{\sc Radio Luminosity Function for the CfA sample}\\
\vspace{0.3cm}
\footnotesize
\begin{tabular}{ccccccc}\hline\hline
\mc{1}{c}{}&
\mc{2}{c}{Seyfert 1s}&
\mc{2}{c}{Seyfert 2s}&
\mc{2}{c}{Both Types}\\ \cline{2-3}\cline{4-5}\cline{6-7}
\mc{1}{c}{$\log L_{rad}^a$}&
\mc{1}{c}{$\log\Phi$}&
\mc{1}{c}{}&
\mc{1}{c}{$\log\Phi$}&
\mc{1}{c}{}&
\mc{1}{c}{$\log\Phi$}&
\mc{1}{c}{}\\
\mc{1}{c}{(ergs~s$^{-1}$)}&
\mc{1}{c}{(Mpc$^{-3}$ M$^{-1}$)}&
\mc{1}{c}{$N$}&
\mc{1}{c}{(Mpc$^{-3}$ M$^{-1}$)}&
\mc{1}{c}{$N$}&
\mc{1}{c}{(Mpc$^{-3}$ M$^{-1}$)}&
\mc{1}{c}{$N$}\\
\hline
 36.00 & $-$4.08 &   1 &  ...    & ... & $-$4.09 &   1\\ 
 36.40 & $-$4.22 &   1 &  ...    & ... & $-$4.23 &   1\\ 
 36.80 & $-$4.90 &   1 &  ...    & ... & $-$4.91 &   1\\ 
 37.20 & $-$4.33 &   2 & $-$4.97 &   1 & $-$4.25 &   3\\ 
 37.60 & $-$4.87 &   5 & $-$4.39 &   5 & $-$4.25 &  10\\ 
 38.00 & $-$4.90 &   2 &  ...    & ... & $-$4.91 &   2\\ 
 38.40 & $-$5.03 &  10 & $-$4.87 &   5 & $-$4.63 &  15\\ 
 38.80 & $-$5.35 &   6 & $-$5.68 &   1 & $-$5.18 &   7\\ 
 39.20 & $-$5.98 &   2 & $-$5.73 &   3 & $-$5.53 &   5\\ 
 39.60 &  ...    & ... & $-$6.73 &   1 & $-$6.72 &   1\\ 
 40.80 & $-$6.79 &   1 &  ...    & ... & $-$6.80 &   1\\ 
\hline
\end{tabular}
\renewcommand{\baselinestretch}{1}
\parbox[c]{12cm}{\noindent\footnotesize
\hspace*{.3cm}$^a$~Central luminosity of a bin 0.4 units wide in $\log
L$, which is equivalent to a width of 1 magnitude.}
\end{center}
\end{table}

\clearpage \twocolumn

\clearpage 

\centerline{\bf FIGURE LEGENDS}

\noindent {\bf Figure 1} --- Radio spectral slope from 6---20~cm versus the
1.5---6~cm slope. Solid line indicates where \al{1.5}{6}~=~\al{6}{20}. The box
encloses those points to the bottom left with both slopes steeper than --0.6.
In this and all plots following (unless otherwise specified) filled symbols are
Seyfert~1s (including 1.0, 1.5, 1.8, and 1.9) and open symbols are
Seyfert~2s. For both Seyfert~1s and~2s: square = 12\um\ sample only; triangle =
CfA sample only; circle = in both the 12\um\ and CfA samples.  On this and all
following plots (except the luminosity functions), the single errorbar shown
represents a typical intrinsic mesurement uncertainty for the parameters
plotted.

\noindent {\bf Figure 2} --- Diagram showing the difference between the
6---20~cm slope derived from single--dish measurements and the same slope as
derived from our VLA data, versus the VLA slopes.

\noindent {\bf Figure 3} --- 6~cm luminosity versus 6~cm compactness
parameter. The straight solid line is the best fit to all the data points, and
the dashed line is the best fit to all except the 7 most luminous (labeled)
points. The curved lines represent models calculated for different values of
the extended luminosity at 6~cm (see text).

\noindent {\bf Figure 4} --- Radio (6~cm) versus infrared (60\um) monochromatic
luminosities. The dotted lines represent various fits to this relation, for
this and other data sets, with the slope constrained to~1. Solid lines
represent calculated models (see text for model parameters).

\noindent {\bf Figure 5} --- IRAS 25---60\um\ slope versus 6~cm compactness
parameter. Horizontal dashed lines represent estimated central and extended
values of the IRAS slope. Curves are calculated models: solid curves represent
lower values of central 6~cm---60\um\ flux ratios and dotted lines represent
higher values.

\noindent {\bf Figure 6} --- Radio Luminosity Function for all objects combined
(Seyfert~1s and~2s) in both the 12\um\ sample (filled circles; offset slightly
for clarity) and the CfA sample (open circles), and in Ulvestad \& Wilson
(1989; x's). Errorbars represent the 90\% confidence interval, based on Poisson
statistics, accurate for very small numbers of data points.  Points with no
error bars represent just one object in that bin.  x's represent the RLF from
Ulvestad \& Wilson (1989).

\noindent {\bf Figure 7} --- Radio Luminosity Function for individual Seyfert
classes in the 12\um\ sample.  Filled squares are Sy~1s and open squares are
Sy~2s.

\noindent {\bf Figure 8} --- Radio Luminosity Function for individual Seyfert
classes in the CfA sample.  Filled triangles are Sy~1s and open triangles are
Sy~2s.


\noindent {\bf Figure 9} --- The ratio of our tapered 6~cm flux to the 4.85~GHz
flux from Becker et al. (1991) versus our 6~cm compactness parameter, R, for
all objects with detections in each work. The dotted lines represent values of
1 for either ratio.

\noindent {\bf Figure 10} --- 2295~MHz flux from Roy et al. versus our 6~cm
compactness parameter. Objects detected at 2295~MHz and with R$<0.85$ are
individually labeled.

\end{document}